\documentclass[twocolumn]{jpsj2} 
\usepackage{graphicx}

\def\runauthor{T. {\sc Maehira} {\it et al.}}

\title{Electronic Properties of Transuranium Compounds \\
with HoCoGa$_5$-Type Tetragonal Crystal Structure}

\author{Takahiro {\sc Maehira}$^{1}$, Takashi {\sc Hotta}$^{2}$,
Kazuo {\sc Ueda}$^3$, and Akira {\sc Hasegawa}$^4$}

\inst{
$^1$Faculty of Science, University of The Ryukyus,
Nishihara, Okinawa 903-0213 \\
$^2$Advanced Science Research Center,
Japan Atomic Energy Research Institute,
Tokai, Ibaraki 319-1195 \\
$^3$Institute for Solid State Physics,
University of Tokyo, 5-1-5 Kashiwa-no-ha, Kashiwa, Chiba 277-8581 \\
$^4$Niigata University, Niigata 950-2181 \\
}

\recdate{\today}

\abst{
By using a relativistic linear augmented-plane-wave method with
the one-electron potential in the local-density approximation,
we investigate energy band structures and the Fermi surfaces
of transuranium compounds NpTGa$_5$, PuTGa$_5$, and AmCoGa$_5$
with transition metal atoms T.
It is found in common that the energy bands in the vicinity of
the Fermi level are mainly due to the large hybridization between
$5f$ and Ga $4p$ electrons.
For PuTGa$_5$, we observe several cylindrical sheets of Fermi surfaces
with large volume for T=Co, Rh, and Ir.
The de Haas-van Alphen (dHvA) frequencies are theoretically estimated
for PuCoGa$_5$.
It is also found that the Fermi surfaces of NpFeGa$_5$, NpCoGa$_5$, and
NpNiGa$_5$ are similar to those of UCoGa$_5$, UNiGa$_5$,
and PuCoGa$_5$, respectively, except for small details.
For AmCoGa$_5$, the Fermi surfaces are found to consist of
large cylindrical electron sheets and small closed hole sheets,
similar to PuCoGa$_5$.
The similarity is basically understood
by the change of electron numbers inside the Fermi surfaces
on the basis of a rigid-band picture.
We discuss our theoretical Fermi surfaces
with the dHvA experimental results on NpTGa$_5$.
}

\kword{Relativistic linear APW method, Fermi surface,
Neptunium compounds, Plutonium compounds, Americium compound}

\begin{document}
\maketitle

%
%
\section{Introduction}

Recently electronic properties of actinide compounds have attracted
renewed attention in the research field of condensed matter physics.
Among numerous kinds of actinide compounds, the group of materials
with HoCoGa$_5$-type tetragonal crystal structure, frequently referred
to as ``115'', has been intensively investigated both from experimental
and theoretical sides.
A characteristic trend in the recent investigation of actinide compounds
has been a rapid expansion of the research frontier from uranium to
transuranium materials, typically found in a series of 115
systems, U-, Np-, Pu-, and Am-115.

First let us briefly survey the properties of U-115.
For several transition metal ions T, UTGa$_5$ are antiferromagnetic
(AF) metals or Pauli paramagnets.
\cite{U115-1a,U115-1b,U115-1c,U115-1d,U115-2,U115-3,U115-4,U115-5,
U115-6,U115-7,U115-8,U115-9}
Among them, neutron scattering experiments have revealed that
UNiGa$_5$ exhibits the G-type AF phase, while UPdGa$_5$ and UPtGa$_5$
have the A-type AF state.\cite{U115-5,U115-9}
Note that G-type indicates a three-dimensional N\'eel state,
while A-type denotes a layered AF structure in which spins align
ferromagnetically in the $ab$ plane and AF along the $c$ axis.
On the other hand, for T=Co, Rh, Ir, Fe, Ru, and Os,
magnetic susceptibility is almost independent of temperature,
since these are Pauli paramagnets.
It is quite interesting that the magnetic structure is different
for U-115 compounds which differ only by the substitution
of transition metal ions.

Now we move on to transuranium 115 compounds.
The rapid progress in the research from uranium to transuranium
systems has been triggered by the discovery of superconductivity
of Pu-115 compound PuCoGa$_5$.\cite{Sarrao,SarraoASR}
It has been reported that the superconducting transition
temperature $T_{\rm c}$ of PuCoGa$_5$ is 18.5K,
which is amazingly high value even compared with
other well-known intermetallic compounds.
The coefficient of electronic specific heat $\gamma$ is estimated
as $\gamma$=77mJ/mol$\cdot$K$^2$, moderately enhanced relative to
that for normal metals, suggesting that PuCoGa$_5$ should be
heavy-fermion superconductor.
In PuRhGa$_5$, superconductivity has been also found.\cite{Wastin}
Although the value of $T_{\rm c}$=8.7K is lower than that of
PuCoGa$_5$, it is still high enough compared with other heavy-fermion
superconductors.
Another Pu-115 material, PuIrGa$_5$, has been also synthesized,
but at least up to now, superconductivity has not been found.
\cite{Wastin2}
In a bulk sample, AF magnetic order has been suggested
at low temperatures, while the experimental results have not yet
converged due to a problem of sample quality.

Recently Np-115 compounds NpTGa$_5$ (T=Fe, Co, and Ni) have been
synthesized and several kinds of physical quantities have been
successfully measured.
\cite{Colineau,Aoki-Ni,Aoki-Co1,Aoki-Co2,Aoki-Rh,Honda,Yamamoto,Homma,Metoki}
Especially, the de Haas-van Alphen (dHvA) effect has been observed
in NpNiGa$_5$ \cite{Aoki-Ni}, which is the first observation of
dHvA signal in transuranium compounds.
For NpCoGa$_5$ and NpRhGa$_5$, the dHvA oscillations have been also detected
and plural number of cylindrical Fermi surfaces are found.
\cite{Aoki-Co2,Aoki-Rh}
For NpFeGa$_5$, the magnetic moment at Fe site has been suggested
in neutron scattering experiments \cite{Metoki} and it has been
also detected by $^{57}$Fe M\"ossbauer spectroscopy.\cite{Homma}
The magnetic structure of Np-115 compounds
also depends sensitively on transition metal ion \cite{Honda,Metoki}:
G-AF for NpNiGa$_5$, A-AF for NpCoGa$_5$, and C-AF for NpFeGa$_5$.
Here C-type indicates a situation in which the ferromagnetic chains
along the $c$ axis are antiferromagnetically coupled in the $ab$ plane.
Note also that in the neutron scattering experiment for NpNiGa$_5$,
the G-AF peak due to canted magnetic moments of Np ions
grows after the FM transition occurs.\cite{Honda,Metoki}
It is characteristic of U-115 and Np-115 compounds that
the magnetic properties are sensitive to the choice of
transition metal ions.

Quite recently, the experimental research on actinide 115 systems
has been further developed and the frontier has reached
americium compound AmCoGa$_5$.\cite{Wastin3}
Interestingly enough, from the resistivity measurement,
superconductivity has been suggested to occur below 2K, but
unfortunately, it has not been confirmed yet by other experimental
techniques, mainly due to the difficulty caused by self-heating effect.
It may be interesting to consider a possibility of superconductivity
in Am-115 also from the theoretical viewpoint.

In order to elucidate the mechanism of superconductivity and
magnetism of such actinide compounds, it is necessary to develop
a microscopic theory based on an appropriate $f$-electron model.
A prescription to construct a microscopic $f$-electron model has
been obtained on the basis of a $j$-$j$ coupling scheme.\cite{Hotta1}
The spirit of the prescription is that the many-body effect
is further included into the band-structure calculation results.
In actual calculations, since the energy dispersion obtained
by the band-structure calculations are very complicated,
the $f$-electron kinetic term is simply reconstructed by using
a tight-binding approximation based on the $j$-$j$ coupling scheme
so as to reproduce the band structure around the Fermi energy.
Here we note that all $5f$ electrons are assumed to be itinerant.
Note also that it is quite natural to use the $j$-$j$ coupling scheme
for the construction of the microscopic $f$-electron model,
since the total angular momentum $j$ is the label for one-electron
state in the relativistic band-structure calculations
in which the Dirac equations are directly solved.

By applying the prescription to uranium compound such as U-115,
it has been clarified that electronic structure of uranium
compounds is effectively described by a two-orbital Hubbard model
based on the $j$-$j$ coupling scheme.\cite{Hotta2}
With increasing $f$-electron number, it has been pointed out
that this two-orbital Hubbard model can be also applied
to some neptunium and plutonium compounds.
For instance, quite recently, magnetic structure of Np-115 \cite{Onishi}
and octupole ordering of NpO$_2$ \cite{Kubo} have been discussed
based on the two-orbital Hubbard model.

Although it is important to develop the microscopic analysis of
actinide compounds based on the simple $f$-electron model,
close attention should be always paid to correct information about
the electronic properties around the Fermi energy obtained by
the relativistic band-structure calculations.
Those two types of researches, microscopic analysis of the simple
$f$-electron model and band-structure calculations, should be
complementary to each other in order to make significant progress
in our understandings on magnetism and superconductivity
of actinide compounds.

In this paper, we study electronic properties of neptunium, plutonium,
and americium compounds such as NpTGa$_5$, PuTGa$_5$, and AmCoGa$_5$
by using a relativistic linear augmented-plane-wave (RLAPW) method with
the one-electron potential in the local-density approximation (LDA).
We observe that the energy bands in the vicinity of
Fermi level are given by the large hybridization between
$5f$ and Ga $4p$ electrons in actinide 115 compounds.
For PuTGa$_5$, several cylindrical sheets of Fermi surfaces
with large volume are found in common for T=Co, Rh, and Ir.
For Np-115, we find that the Fermi surfaces of NpFeGa$_5$, NpCoGa$_5$,
and NpNiGa$_5$ are similar to those of UCoGa$_5$, UNiGa$_5$,
and PuCoGa$_5$, respectively.
It is also found that the Fermi surfaces of AmCoGa$_5$ are rather
similar to those of PuCoGa$_5$.
Basically, the similarity is understood by the rigid-band nature of
the 115 structure, since the topology of the Fermi surfaces
in the band-structure calculations seems to depend on the sum of
$f$-electron number of actinide ions and $d$-electron number of
transition metal ions.

The organization of this paper is as follows.
In Sec.~2, we briefly explain the method of our band-structure
calculations and lattice parameters for PuTGa$_5$, NpTGa$_5$,
and AmCoGa$_5$.
In Sec.~3, we show the results for electronic band structure and
the Fermi surfaces of PuTGa$_5$ (T=Co, Rh, and Ir),
NpTGa$_5$ (T=Fe, Co, and Ni), and AmCoGa$_5$.
In Sec.~4, we will discuss our present results of Pu-, Np-, and Am-115
materials in comparison with that of UCoGa$_5$.
We also discuss the dHvA results on NpTGa$_5$ from the band-theoretical
viewpoint. Finally, in Sec.~5, we will summarize this paper.

%
%
\section{Method of Band Calculation}

In order to calculate the electronic energy band structure of
transuranium compounds, in general,
it is necessary to include relativistic effects such as
relativistic energy shifts, relativistic screening effects,
and spin-orbit interaction.\cite{Hasegawa}
Based on the Dirac equation, such relativistic effects
can be fully taken into account in the energy band structure
calculation.\cite{Loucks,Yamagami1,Andersen}
Among several kinds of the band-structure calculation techniques,
in this paper we adopt the RLAPW method.\cite{RLAPW}
The exchange and correlation potential is considered within the LDA,
while the spatial shape of the one-electron potential is determined
in the muffin-tin approximation.
The self-consistent calculation is performed by using the
lattice constants which are determined experimentally.
It is stressed that all $5f$ electrons in PuTGa$_5$, NpTGa$_5$,
and AmCoGa$_5$
are assumed to be itinerant in our band-structure calculations.

\begin{figure}[t]
\includegraphics[width=0.6\linewidth]{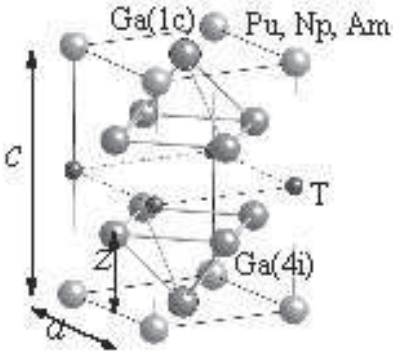}
\caption{Crystal structure of PuTGa$_5$, NpTGa$_5$ and AmCoGa$_5$.}
\label{fi1}
\end{figure}

In Fig.~\ref{fi1}, we show the HoCoGa$_5$-type tetragonal crystal
structure which belongs to the space group P4/mmm (No.~123)
and ${\rm D_{4h}^{1}}$.
Note that one molecule is contained per primitive cell.
In the HoCoGa$_5$-type tetragonal structure,
positions of atoms in the unit cell are given by
(0, 0, 0) for actinide atom,
(0, 0, 1/2) for transition metal atom,
(1/2, 1/2, 0) for Ga atom at $1c$ site, and
(0, 1/2, $\pm z$) for Ga atom at $4i$ site,
where $z$ is a parameter determined from X-ray diffraction
experiments and the labels for atoms are referred in Fig.~\ref{fi1}.
The lattice constants, $a$ and $c$, and the $z$ parameters
for NpTGa$_5$, PuTGa$_5$, and AmCoGa$_5$ are listed in
Tables \ref{ta2} and \ref{ta1}.

\begin{table}[t]
\caption{Lattice constants and $z$ parameter of NpFeGa$_5$,
NpCoGa$_5$ and NpNiGa$_5$ determined experimentally.
\cite{Aoki-Ni,Aoki-Co1}}
\label{ta2}
\begin{tabular}{llll}
    & NpFeGa$_5$ & NpCoGa$_5$ & NpNiGa$_5$ \\
$a$ & 4.257 \AA & 4.237 \AA & 4.231 \AA \\
$c$ & 6.761 \AA & 6.787 \AA & 6.783 \AA \\
$z$ & 0.3011 & 0.310 & 0.3135
\end{tabular}
\end{table}

\begin{table}[t]
\caption{Lattice constants and $z$ parameters of PuCoGa$_5$,
PuRhGa$_5$, PuIrGa$_5$, and AmCoGa$_5$ determined experimentally.
\cite{Sarrao,Wastin,Wastin2,Wastin3}}
\label{ta1}
\begin{tabular}{lllll}
    & PuCoGa$_5$ & PuRhGa$_5$ & PuIrGa$_5$ & AmCoGa$_5$ \\
$a$ & 4.232 \AA & 4.301 \AA & 4.324 \AA & 4.233 \AA \\
$c$ & 6.786 \AA & 6.856 \AA & 6.817 \AA & 6.823 \AA \\
$z$ & 0.312 & 0.306 & 0.302 & 0.3106
\end{tabular}
\end{table}

\begin{table}[t]
\caption{The spin-orbit splitting obtained in the relativistic
atomic calculation. Energy unit is milli-Ryd. and 1 Ryd.=13.6 eV.}
\vspace{5mm}
\label{ta-so}
\begin{tabular}{|c|c|c|c|c|c|} \hline
Pu $5f$ & Pu $6d$ & Np $5f$ & Np $6d$ & Am $5f$ & Am $6d$ \\ \hline
80 & 37 & 70 & 38 & 90 & 36 \\ \hline
\hline
Fe $3d$ & Co $3d$ & Ni $3d$ & Rh $4d$ & Ir $5d$ & Ga $4p$ \\ \hline
11 & 13 & 18 & 29 & 90 & 7 \\ \hline
\end{tabular}
\end{table}


The iteration process for solving the Dirac one-electron equation starts
with the crystal charge density that is constructed by
superposing the relativistic atomic charge densities for neutral atoms
Pu([Rn]$5f^{5}$$6d^{1}$$7s^{2}$), Np([Rn]$5f^{4}$$6d^{1}$$7s^{2}$), 
Am([Rn]$5f^{6}$$6d^{1}$$7s^{2}$), 
Fe([Ar]$3d^{6}$$4s^{2}$), Co([Ar]$3d^{7}$$4s^{2}$),
Ni([Ar]$3d^{8}$$4s^{2}$),
Rh([Kr]$4d^{8}$$5s^{1}$), Ir([Xe]$4f^{14}$$5d^{7}$$6s^{2}$),
and Ga([Ar]$3d^{10}$$4p^{1}$$4s^{2}$),
where [Rn], [Kr], [Xe], and [Ar] symbolically indicate the
closed electronic configuration for radon, krypton, xenon,
and argon, respectively.
In the calculation for the atoms, the same exchange and correlation
potential are used as for the crystal.
We assume that the Rn core state except the $6p^{6}$ state for Pu, Np,
and Am, the Kr core state for Rh, the Xe core state for Ir,
the Ar core state for Co, and
the Ar core state for Ga are unchanged during the iteration.
Namely, the frozen-core approximation is adopted for these core states
in the calculation for the crystal.
The values of the spin-orbit splitting in the relativistic atomic
calculation are listed in Table \ref{ta-so}.

In each iteration step for the self-consistent calculation processes,
a new crystal charge density is constructed using eighteen $k$ points,
which are uniformly distributed in the irreducible 1/16 part of
the Brillouin zone.
At each $k$ in the Brillouin zone, 431 plane waves are adopted
under the condition $|k+G|$$\leq$$4.4(2\pi/a)$
with $G$ the reciprocal lattice vector and angular momentum up to
$\ell_{\rm max}$=8 are taken into account.

%
%
\section{Band Calculation Results}

\subsection{Results for Pu-115}

First let us discuss the calculated results for Pu-115 compounds,
since PuCoGa$_5$ and PuRhGa$_5$ have been recently focused as
``high-$T_{\rm c}$'' $f$-electron superconductors.
Note that a part of the results of PuCoGa$_5$ has been published.
\cite{Maehira1}
Band-structure calculation results for PuTGa$_5$ (T=Co, Rh, and Ir)
were also reported by another group.\cite{Opahle1,Opahle2}

\begin{figure}[t]
\includegraphics[width=1.0\linewidth]{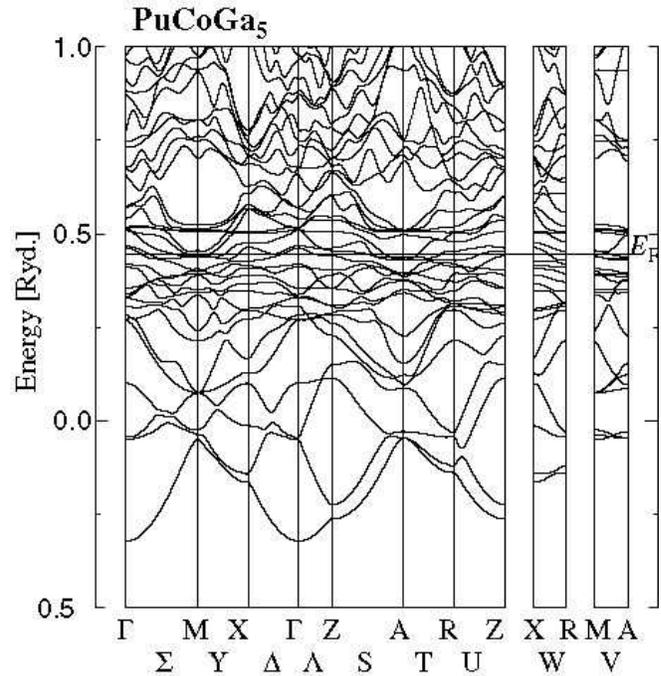}
\caption{Energy band structure for PuCoGa$_5$ calculated by
using the self-consistent RLAPW method.
$E_{\rm F}$ indicate the position of the Fermi level.}
\label{fi2}
\end{figure}

\begin{figure}[t]
\includegraphics[width=1.0\linewidth]{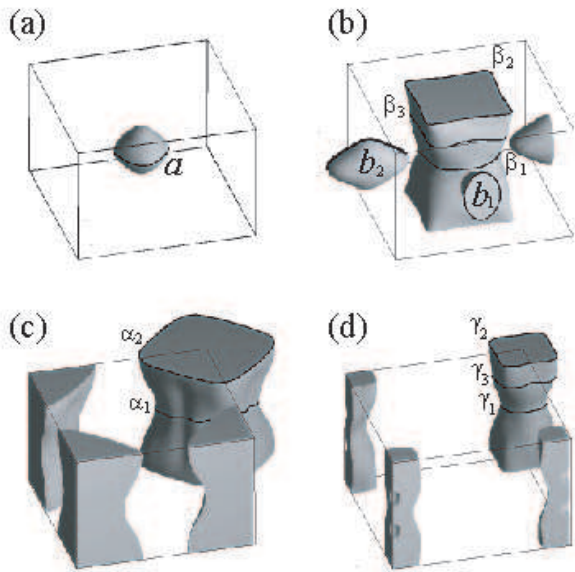}
\caption{Calculated Fermi surfaces of PuCoGa$_5$ for
(a) 15th band hole sheets, (b) 16th band hole sheets, 
(c) 17th band electron sheets, and (d) 18th band electron sheets.
The center of the Brillouin zone is set at the $\Gamma$ point. We also
show various kinds of extremal cross section of the Fermi surface.}
\label{fi3}
\end{figure}

\subsubsection{PuCoGa$_5$}

In Fig.~\ref{fi2}, we depict the energy band structure
along the symmetry axes in the Brillouin zone
in the energy region from $-0.5$ Ryd. to 1.0 Ryd.
Note here that the three Pu $6p$ and twenty-five Ga $3d$ bands
in the energy range between $-1.0$ Ryd. and $-0.6$ Ryd. are not shown
in Fig.~\ref{fi2}, since those bands are irrelevant to the present discussion.
The Fermi level $E_{\rm F}$ is located at 0.446 Ryd. and
in the vicinity of $E_{\rm F}$, there occurs a hybridization between
the Pu 5$f$ and Ga 4$p$ states.
Around $E_{\rm F}$ near M point, the flat 5$f$ bands split into two groups,
corresponding to the total angular momentum $j$=5/2 (lower bands) and
7/2 (upper bands).
The magnitude of the splitting between those groups is estimated
as 1.0 eV, which is almost equal to the spin-orbit splitting
in the atomic $5f$ state.

The number of the valence electrons in the APW sphere is partitioned
into the angular momenta as listed in Table \ref{ta3}.
There are 8.17 valence electrons outside the APW sphere
in the primitive cell.
The total density of states at $E_{\rm F}$ is evaluated as 
$N(E_{\rm F})$=97.3 states/Ryd.cell.
By using this value, the theoretical specific heat coefficient
$\gamma_{\rm band}$ is estimated as 16.9 mJ/K$^2 \cdot$mol.
We note that the experimental electronic specific heat coefficient 
$\gamma_{\rm exp}$ is 77 mJ/K$^2 \cdot$mol.\cite{Sarrao}
We define the enhancement factor for the electronic specific heat
coefficient as $\lambda$=$\gamma_{\rm exp}/\gamma_{\rm band}$$-$1,
and in the present case, we obtain $\lambda$=3.6.
The disagreement between $\gamma_{\rm band}$ and $\gamma_{\rm exp}$
values is ascribed to electron correlation effect and electron-phonon
interactions, which are not fully taken into account in the present
LDA band theory.

\begin{table}[t]
\caption{The number of valence electrons for PuCoGa$_5$
in the Pu APW sphere,
the Co APW sphere, and the Ga APW sphere partitioned into
angular momenta.}
\label{ta3}
\begin{tabular}{lrrrr}
\multicolumn{1}{c}{ } &
\multicolumn{1}{c}{$s$} &
\multicolumn{1}{c}{$p$} &
\multicolumn{1}{c}{$d$} &
\multicolumn{1}{c}{$f$} \\ \hline
Pu & 0.39 & 6.15 & 1.81 & 5.24 \\
Co & 0.43 & 0.44 & 7.46 & 0.01 \\
Ga($1c$) & 0.95 & 0.69 & 9.92 & 0.01 \\
Ga($4i$) & 3.73 & 2.82 & 39.75 & 0.06 \\
\end{tabular}
\end{table}

Now let us discuss the Fermi surfaces of PuCoGa$_5$.
In Fig.~\ref{fi2}, the lowest fourteen bands are fully occupied.
The next four bands are partially occupied, while higher bands are
empty. Namely, 15th, 16th, 17th, and 18th bands crossing
the Fermi level construct the hole or electron sheet of the Fermi
surface, as shown in Fig.~\ref{fi3}.
The Fermi surface from the 15th band consists of
one hole sheet centered at the $\Gamma$ point.
The 16th band has two kinds of sheets, as shown in Fig.~\ref{fi3}(b).
One is a set of small hole pockets, each of which is centered
at the X point.
Another is a large cylindrical hole sheet
which is centered at the $\Gamma$ point.
The 17th band forms a large cylindrical electron sheet
which is centered at the M point.
The 18th band has a slender cylindrical electron sheet
which is also centered at the M point.
These electron sheets are characterized by two-dimensional
Fermi surfaces.
The number of carriers contained in these Fermi-surface sheets are
0.040 holes/cell, 0.563 holes/cell, 0.519 electrons/cell, and
0.084 electrons/cell in the 15th, 16th, 17th, and 18th bands,
respectively.
The total number of holes is equal to that of electrons,
which represents that ${\rm PuCoGa_{5}}$ is a compensated metal.

Note that the Fermi surfaces of PuTGa$_{5}$
are similar to those of CeTIn$_{5}$.\cite{Maehira2}
This similarity can be understood based on the electron-hole
conversion relation in the $j$-$j$ coupling scheme,
\cite{Maehira1}
since one $f$ electron is included in the $j$=5/2 sextet
for Ce$^{3+}$ ion,
while five $f$ electrons are contained for Pu$^{3+}$ ion.

\begin{figure}[t]
\includegraphics[width=1.0\linewidth]{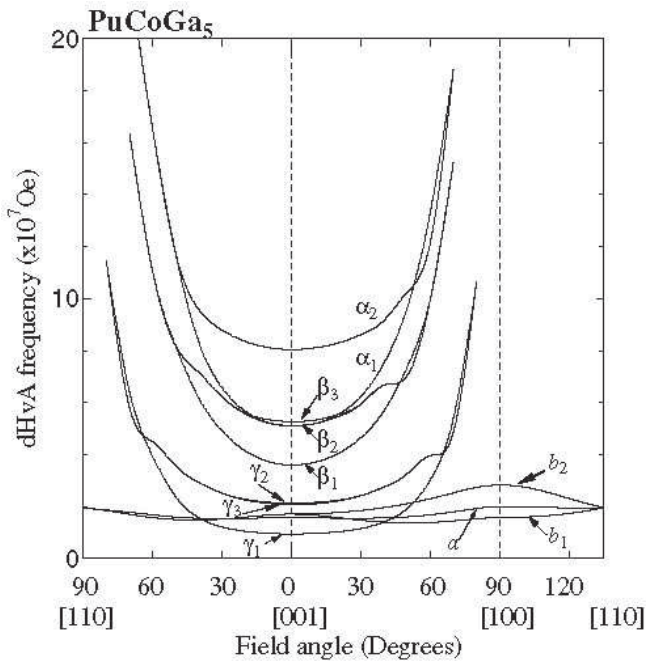}
\caption{Angular dependence of the theoretical dHvA frequency in
PuCoGa$_5$. The labels of the branches are referred in Fig.~\ref{fi3}.}
\label{fi4}
\end{figure}

In Fig.~\ref{fi3}, various kinds of extremal orbits are
depicted on the Fermi surfaces.
Among them, $a$ denotes an orbit running around the small closed
hole sheet centered  at the $\Gamma$ point in the 15th band,
while $b_1$ and $b_2$ indicate a couple of dHvA frequency branches
which originate from the 16th hole sheets and
exist in the narrow range from 0.015 MOe and 0.03 MOe.
Its center is located at the X point.
On the other hand, $\alpha_{1}$ and $\alpha_{2}$ are the orbits
running around the cylinder along the V axis in the 17th band,
while $\beta_{1}$, $\beta_{2}$, and $\beta_{3}$ are the orbits
running around the cylinder along the $\Lambda$ axis in the 16th band.
Finally, $\gamma_{1}$, $\gamma_{2}$, and $\gamma_{3}$ denote
the orbits running around the cylinder along the V axis in the 18th band.
These orbits exist in the large range of angles in the vicinity of
the [001] direction.

In Fig.~\ref{fi4}, we show the angular dependence of the theoretical
dHvA frequency in PuCoGa$_5$.
The area of the extremal cross section of the Fermi surface $A$
is related to the dHvA frequency $F$ by the well-known formula
$F$=$(c \hbar / 2 \pi)A$.
The Fermi surface produces many dHvA frequencies in the wide frequency
range between 0.01 MOe and 220 MOe, as shown in Fig.~\ref{fi4}.
The small hole sheet in the 15th band possesses dHvA frequencies
in the range between 0.015 MOe and 0.02 MOe.
The Fermi surface in the 16th, 17th, and 18th band possesses
many extremal cross section in the limited range of angles
because of its rugged shape.
The dHvA experiments on Pu-115 compounds have not yet been
carried out due to several experimental difficulties, but
we believe that such experiments will be done in near future.
On that occasion, the present theoretical results will be
helpful for the experimentalists.

\begin{figure}[t]
\includegraphics[width=1.0\linewidth]{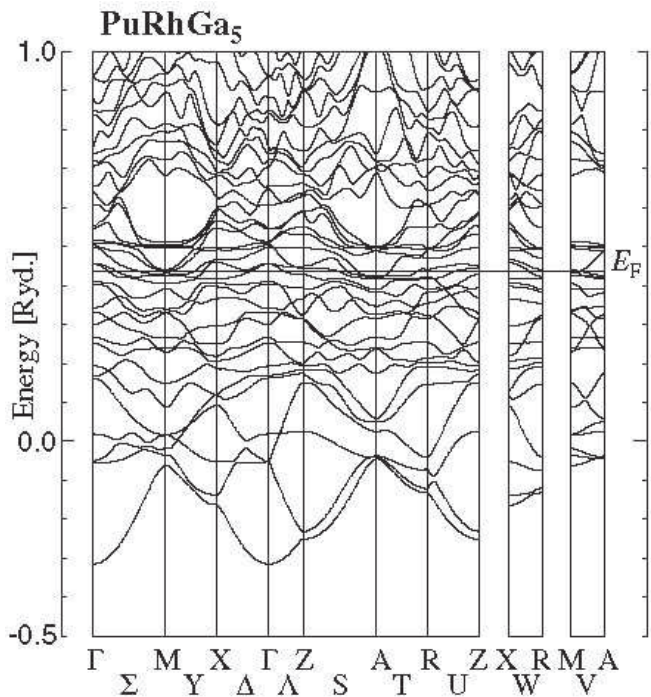}
\caption{Energy band structure for PuRhGa$_5$ calculated by
using the self-consistent RLAPW method.}
\label{fi5}
\end{figure}

\begin{table}[t]
\caption{The number of valence electrons for PuRhGa$_5$
in the Pu APW sphere, the Rh APW sphere,
and the Ga APW sphere partitioned into angular momenta.}
\label{ta4}
\begin{tabular}{lrrrr}
\multicolumn{1}{c}{ } & 
\multicolumn{1}{c}{$s$} & 
\multicolumn{1}{c}{$p$} & 
\multicolumn{1}{c}{$d$} & 
\multicolumn{1}{c}{$f$} \\ \hline
Pu & 0.33 & 6.16 & 1.85 & 5.22 \\
Rh & 0.35 & 0.35 & 7.38 & 0.02 \\
Ga($1c$) & 0.99 & 0.75 & 9.94 & 0.01 \\
Ga($4i$) & 3.73 & 2.84 & 39.77 & 0.07 \\
\end{tabular}
\end{table}

\begin{figure}[t]
\includegraphics[width=1.0\linewidth]{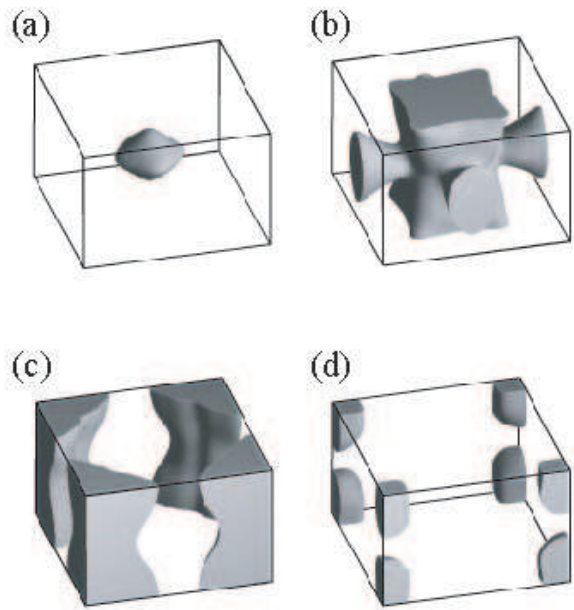}
\caption{Calculated Fermi surfaces of PuRhGa$_5$ for
(a) 15th band hole sheets, (b) 16th band hole sheets,
(c) 17th band electron sheets, and (d) 18th band electron sheets.}
\label{fi6}
\end{figure}

\subsubsection{PuRhGa$_5$}

Let us move to another superconducting Pu-115 material PuRhGa$_5$.
In Fig.~\ref{fi5}, we depict the energy band structure and
the Fermi energy $E_{\rm F}$ is located at 0.436 Ryd.
Narrow Pu $5f$-bands just above $E_{\rm F}$ are split into
two subbands by the spin-orbit interaction.
A hybridization between the Pu 5$f$ state and Ga 4$p$ state
occurs in the vicinity of $E_{\rm F}$ also for PuRhGa$_5$.
In Table~\ref{ta4}, we list the number of the valence electrons
in the APW sphere distributed into the angular momenta.
There are 8.31 valence electrons outside the APW sphere
in the primitive cell.
The theoretical electronic specific-heat coefficient
$\gamma_{\rm band}$ is 13.2 mJ/K$^2 \cdot$mol,
which is smaller than that of PuCoGa$_5$.

Since 15th, 16th, 17th, and 18th bands are partially occupied,
these four bands construct the Fermi surface,
as in the case of PuCoGa$_5$.
The hole and electron sheets of the Fermi surface in PuRhGa$_5$
are shown in Fig.~\ref{fi6}.
The Fermi surface from the 15th band consists of one hole sheet
centered at the $\Gamma$ point.
The hole sheet centered at the $\Gamma$ point in the 16th band
is shown in Fig.~\ref{fi6}(b).
This Fermi surface possesses the crossed arms,
which are connected like a jungle-gym to the next Brillouin zone.
The 17th band has a cylindrical electron sheet centered at the M point.
The Fermi surface from the 18th band consists of one hole sheet
centered at the A point.

The number of carriers contained in these Fermi-surface sheets are
0.040 holes/cell, 0.627 holes/cell, 0.594 electrons/cell,
and 0.073 electrons/cell in the 15th, 16th, 17th, and 18th bands,
respectively.
The total number of holes is equal to that of electrons,
since ${\rm PuRhGa_{5}}$ is also a compensated metal.

\begin{figure}[t]
\includegraphics[width=1.0\linewidth]{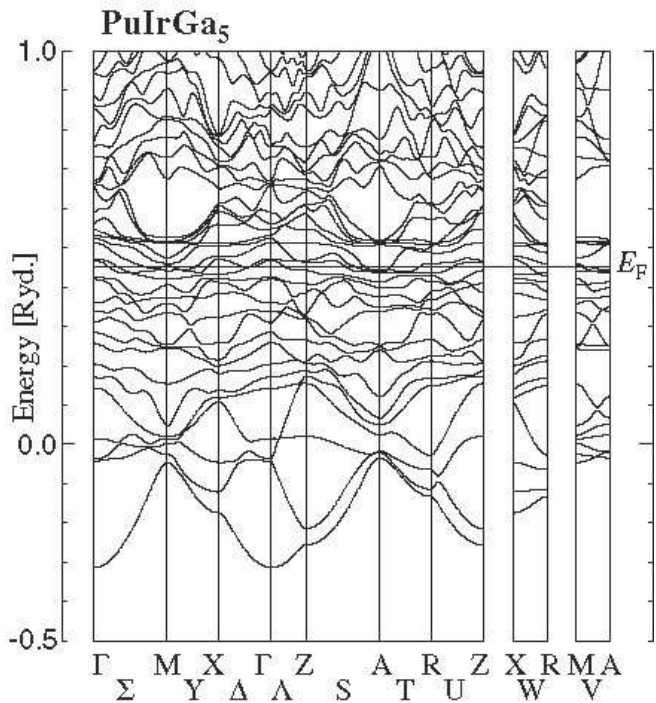}
\caption{Energy band structure for PuIrGa$_5$ calculated by
using the self-consistent RLAPW method.
$E_{\rm F}$ indicate the position of the Fermi level.}
\label{fi7}
\end{figure}

\begin{table}[t]
\caption{
The number of valence electrons for PuIrGa$_5$
in the Pu APW sphere, the Ir APW sphere,
and the Ga APW sphere partitioned into angular momenta.}
\label{ta5}
\begin{tabular}{lrrrr}
\multicolumn{1}{c}{ } & 
\multicolumn{1}{c}{$s$} & 
\multicolumn{1}{c}{$p$} & 
\multicolumn{1}{c}{$d$} & 
\multicolumn{1}{c}{$f$} \\ \hline
Pu & 0.36 & 6.13 & 1.72 & 5.26 \\
Ir & 0.52 & 0.38 & 6.93 & 0.02 \\
Ga($1c$) & 1.01 & 0.78 & 9.96 & 0.01 \\
Ga($4i$) & 3.71 & 2.72 & 39.76 & 0.06 \\
\end{tabular}
\end{table}

\begin{figure}[t]
\includegraphics[width=1.0\linewidth]{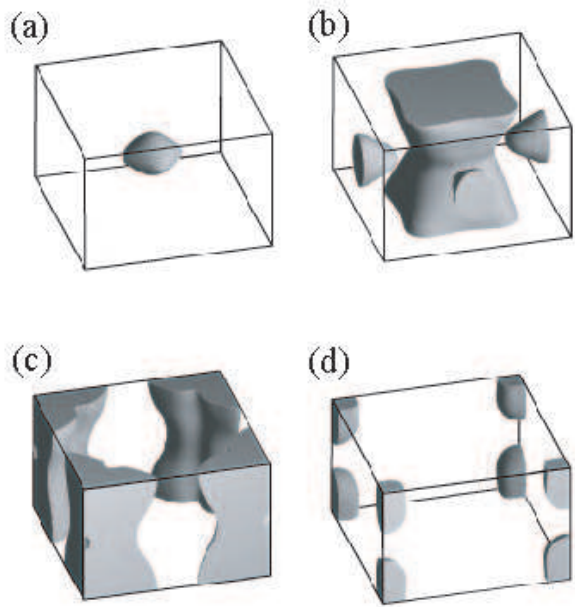}
\caption{Calculated Fermi surfaces of PuIrGa$_5$ for
(a) 15th band hole sheets, (b) 16th band hole sheets, 
(c) 17th band electron sheets, and (d) 18th band electron sheets.}
\label{fi8}
\end{figure}

\subsubsection{PuIrGa$_5$}

Now we discuss the electronic properties of yet another Pu-115 compound.
In Fig.~\ref{fi7}, the energy band structure for PuIrGa$_5$
in the vicinity of the Fermi energy $E_{\rm F}$ is shown
along high-symmetry axes in the Brillouin zone.
The Fermi energy $E_{\rm F}$ is located at 0.453 Ryd.
The band structure around the Fermi energy is almost the same
as those of PuCoGa$_5$ and PuRhGa$_5$.
In Table~\ref{ta5}, we show the number of the valence electrons
in the APW sphere, which are partitioned into the angular momenta.
There are 8.64 valence electrons outside the APW sphere in the
primitive cell.
In each Pu APW sphere, 5.26 electrons with the 5$f$ symmetry are 
contained. The theoretical electronic specific-heat coefficient
$\gamma_{\rm band}$ is 11.5 mJ/K$^2 \cdot$mol,
which is smaller than that of PuRhGa$_5$.

The Fermi level crosses the 15th, 16th, 17th, and 18th bands,
which produce various hole and electron sheets of the Fermi surface.
The hole and electron sheets of the Fermi surface in
PuIrGa$_5$ are shown in Fig.~\ref{fi8}.
Again we see that the Fermi surface structure is quite similar to
that of PuCoGa$_5$.
The Fermi surface from the 15th band consists of
one hole sheet centered at the $\Gamma$ point.
The 16th band has a large cylindrical hole sheet which is centered
at the $\Gamma$ point and two equivalent small hole pockets
at the X point.
The 17th band has a large electron sheet which is open in the
[001] direction and looks like a cylinder running along the V axis.
The Fermi surface from the 18th band consists of
one hole sheet centered at the A point.

The number of carriers contained in these Fermi-surface sheets
are 0.025 holes/cell, 0.579 holes/cell, 0.537 electrons/cell,
and 0.067 electrons/cell in the 15th, 16th, 17th, and 18th bands,
respectively.
The total number of holes is equal to that of electrons,
again indicating that PuIrGa$_5$ is a compensated metal.

Among three Pu-115 compounds, it has been found that
there is no essential differences in the Fermi surface structure,
when Co is substituted by Rh and Ir.
The energy scale of the band-structure calculation is
considered to be too large
to detect the small change in the Fermi surface
structure among three Pu-115 materials.


\subsection{Results for Np-115}

Let us turn our attention to Np-115 compounds.
It is quite impressive that NpTGa$_5$ with T=Fe, Co, Ni, and Rh
have been synthesized and several kinds of physical quantities
have been measured.
Especially, the dHvA signals have been successfully detected
and the comparison with band-calculation results is important.
Note that a part of the results on NpCoGa$_5$ has been reported.
\cite{Maehira1,Opahle2}

\begin{figure}[t]
\includegraphics[width=1.0\linewidth]{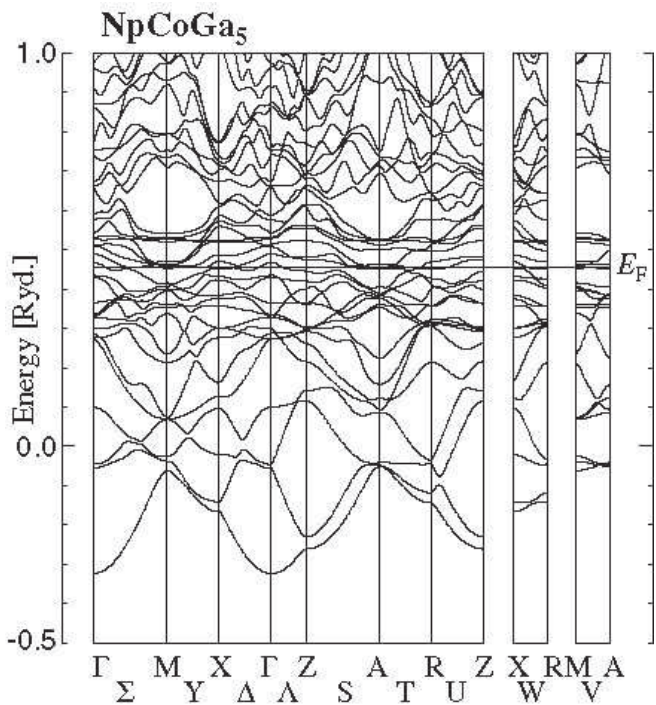}
\caption{Energy band structure for NpCoGa$_5$ calculated by
using the self-consistent RLAPW method.
$E_{\rm F}$ indicate the position of the Fermi level.}
\label{fi9}
\end{figure}

\begin{table}[t]
\caption{The number of valence electrons for NpCoGa$_5$
in the Np APW sphere, the Co APW sphere,
and the Ga APW sphere partitioned into angular momenta.}
\label{ta6}
\begin{tabular}{lrrrr}
\multicolumn{1}{c}{ } & 
\multicolumn{1}{c}{$s$} & 
\multicolumn{1}{c}{$p$} & 
\multicolumn{1}{c}{$d$} & 
\multicolumn{1}{c}{$f$} \\ \hline
Np & 0.32 & 6.11 & 1.91 & 4.17 \\
Co & 0.41 & 0.43 & 7.49 & 0.01 \\
Ga($1c$) & 0.98 & 0.73 & 9.93 & 0.01 \\
Ga($4i$) & 3.75 & 2.90 & 39.75 & 0.06 \\
\end{tabular}
\end{table}

\begin{figure}[t]
\includegraphics[width=1.0\linewidth]{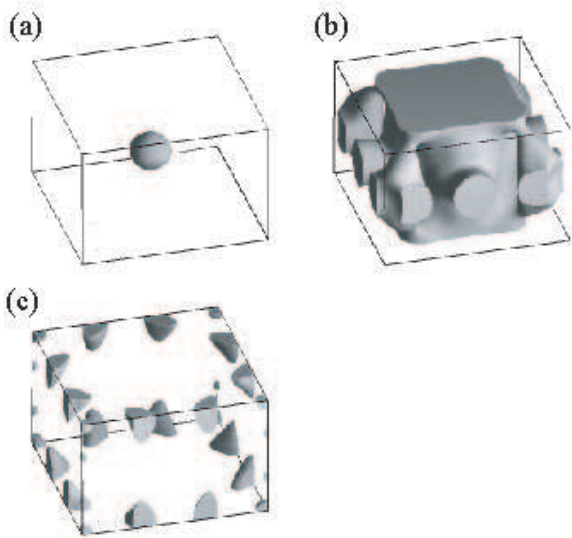}
\caption{Calculated Fermi surfaces of NpCoGa$_5$ for
(a) 15th band hole sheets, (b) 16th band hole sheets, and
(c) 17th band electron sheets.}
\label{fi10}
\end{figure}

\subsubsection{NpCoGa$_5$}

First let us show the results of NpCoGa$_5$.
Figure \ref{fi9} indicates the energy band structure calculated
for NpCoGa$_5$, in the energy range from $-0.5$ Ryd. to 1.0 Ryd.
Note that the three Np $6p$ and twenty-five Ga $3d$ bands
in the energy range between $-1.0$ Ryd. and $-0.6$ Ryd. are
not shown in Fig.~\ref{fi9}.
The Fermi level $E_{\rm F}$ is located at 0.456 Ryd. and
in the vicinity of $E_{\rm F}$, there occurs a hybridization
between the Np 5$f$ and Ga 4$p$ states, similar to the case of
PuTGa$_5$ and UTGa$_5$.
The hybridization between $5f$ and Ga $4p$ electrons is considered
to be a common feature in actinide 115 materials.
In Fig.~\ref{fi9}, the flat bands in the narrow energy range
just above $E_{\rm F}$ consist dominantly of the Np $f$ states.
Above $E_{\rm F}$ near M point, the flat 5$f$ bands split into two
groups, corresponding to the total angular momentum $j$=5/2
(lower bands) and 7/2 (upper bands).
The magnitude of the splitting between those groups is estimated
as 0.95 eV, a little bit smaller than that of Pu,
which is almost equal to the spin-orbit splitting
in the atomic $5f$ state.

The number of the valence electrons in the APW sphere is
listed in Table~\ref{ta6}. 
There are 8.11 valence electrons outside the APW sphere
in the primitive cell and each Np APW sphere contains
about 4.17 electrons in the $f$ state.
The total density of states at $E_{\rm F}$ is evaluated as 
$N(E_{\rm F})$=230.4 states/Ryd.cell.
By using this value, the theoretical specific heat coefficient
$\gamma_{\rm band}$ is estimated as 39.9 mJ/K$^2 \cdot$mol.
Since the experimental electronic specific heat coefficient 
$\gamma_{\rm exp}$ is 60.0 mJ/K$^2 \cdot$mol,\cite{Aoki-Co1}
the enhancement factor $\lambda$ is found to be 0.5.

Next we discuss the Fermi surfaces of NpCoGa$_5$.
In Fig.~\ref{fi9}, the lowest fourteen bands are fully occupied.
The next three bands are partially occupied, while higher bands are
empty, indicating that 15th, 16th, and 17th bands crossing
the Fermi level form the hole or electron sheet of the Fermi surface,
as shown in Fig.~\ref{fi10}.
The Fermi surface from the 15th band consists of
small hole sheets centered at the $\Gamma$ point.
The 16th band constructs a large cylindrical hole sheet centered
at the $\Gamma$ point, which exhibits a complex network
consisting of big ``arms'' along the edges
of the Brillouin zone, as observed in Fig.~\ref{fi10}(b).
The 17th band constructs a small electron sheet,
as shown in Fig.~\ref{fi10}(c).

The number of carriers contained in these Fermi-surface sheets are 
0.015 holes/cell, 0.996 holes/cell, and 0.063 electrons/cell 
in the 15th, 16th, and 17th bands, respectively.
The total number of holes is not equal to that of electrons,
indicating that ${\rm NpCoGa_{5}}$ is an uncompensated metal.

\subsubsection{NpNiGa$_5$}

\begin{figure}[t]
\includegraphics[width=1.0\linewidth]{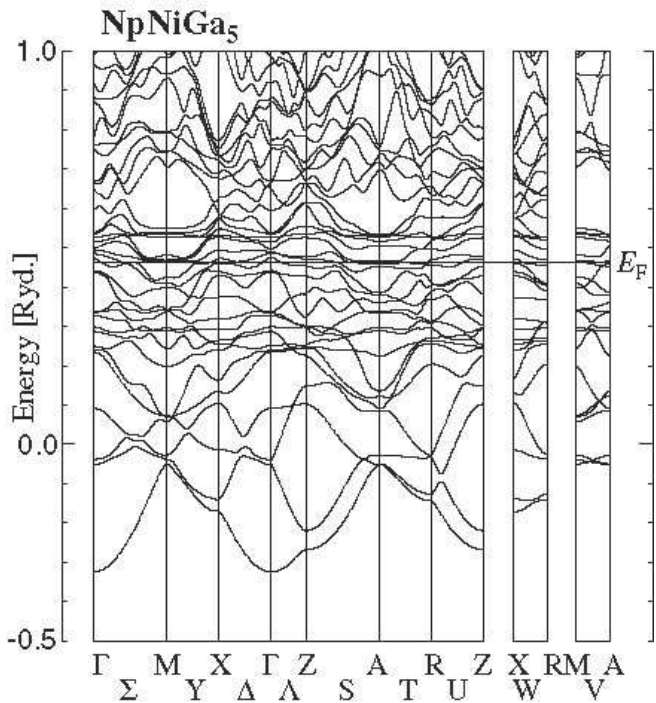}
\caption{Energy band structure calculated for NpNiGa$_5$ with
the self-consistent RLAPW method.}
\label{fi11}
\end{figure}

Next we discuss the electronic properties of NpNiGa$_5$.
In Fig.~\ref{fi11}, the energy-band structure is depicted.
We should note that there is no qualitative difference
in the energy-band structure between NpCoGa$_5$ and NpNiGa$_5$,
but the position of $E_{\rm F}$ is a little bit shifted
as $E_{\rm F}$=0.464 Ryd.
The number of the valence electrons in the APW sphere is
listed for each angular momenta in Table~\ref{ta7}.
There are 8.44 valence electrons outside the APW sphere in the
primitive cell and each Np APW sphere contains about 4.15 electrons
in the $f$ state.
The total density of states is calculated at $E_{\rm F}$ as 
$N(E_{\rm F})=171.9$ states/Ryd.cell, leading to
$\gamma_{\rm band}$=29.8 mJ/K$^2 \cdot$mol.
Note that $\gamma_{\rm exp}$=100.0 mJ/K$^2 \cdot$mol for NpNiGa$_5$,
leading to $\lambda$=2.4.

\begin{figure}[t]
\includegraphics[width=1.0\linewidth]{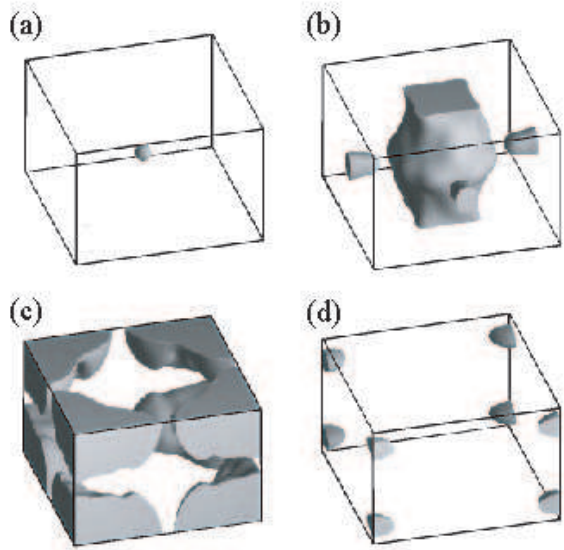}
\caption{Calculated Fermi surfaces of ${\rm NpNiGa_{5}}$ 
for (a) 15th band hole sheets, (b)16th band hole sheets, 
(c)17th band electron sheets and (d)18th band electron sheets.}
\label{fi12}
\end{figure}

In Fig.~\ref{fi12}, we show the Fermi surfaces formed by
15th, 16th, 17th, and 18th bands.
The Fermi surface from the 15th band includes one small hole sheet
centered at the $\Gamma$ point.
The 16th band constructs a large cylindrical hole sheet centered
at the $\Gamma$ point, while two equivalent small hole sheets are
centered at X points.
The 17th band has a large electron sheet centered at the A point.
This large electron sheet looks like a swelled square-cushion
with bulges on surfaces.
The 18th band has a small electron sheet.
Each electron sheet lies across the V axis and looks like a cushion.
If the 16th-band hole Fermi surface of NpCoGa$_5$ is almost occupied
by an electron and the volume of the 17th-band electron Fermi surface
is slightly enlarged, these Fermi surfaces correspond to
the 16th-band hole Fermi surfaces and the 17th-band electron ones
of NpNiGa$_5$, respectively.

The numbers of carriers contained in these Fermi-surface sheets are
0.006 holes/cell, 0.430 holes/cell, 0.427 electrons/cell, and
0.009 electrons/cell in the 15th, 16th, 17th, and 18th bands,
respectively.
The total number of holes is equal to that of electrons,
indicating that NpNiGa$_5$ is a compensated metal.

\begin{table}[t]
\caption{The number of the valence electrons for NpNiGa$_5$
in the Np APW sphere, the Ni APW sphere,
and the Ga APW sphere partitioned into angular momenta.}
\label{ta7}
\begin{tabular}{lrrrr}
\multicolumn{1}{c}{ } & 
\multicolumn{1}{c}{$s$} & 
\multicolumn{1}{c}{$p$} & 
\multicolumn{1}{c}{$d$} & 
\multicolumn{1}{c}{$f$}   \\ \hline
Np & 0.38 & 6.12 & 1.91 & 4.15 \\
Ni & 0.48 & 0.49 & 8.39 & 0.01 \\
Ga($1c$) & 0.94 & 0.67 & 9.92 & 0.01 \\
Ga($4i$) & 3.72 & 2.79 & 39.73 & 0.05 \\
\end{tabular}
\end{table}

\subsubsection{NpFeGa$_5$}

\begin{figure}[t]
\includegraphics[width=1.0\linewidth]{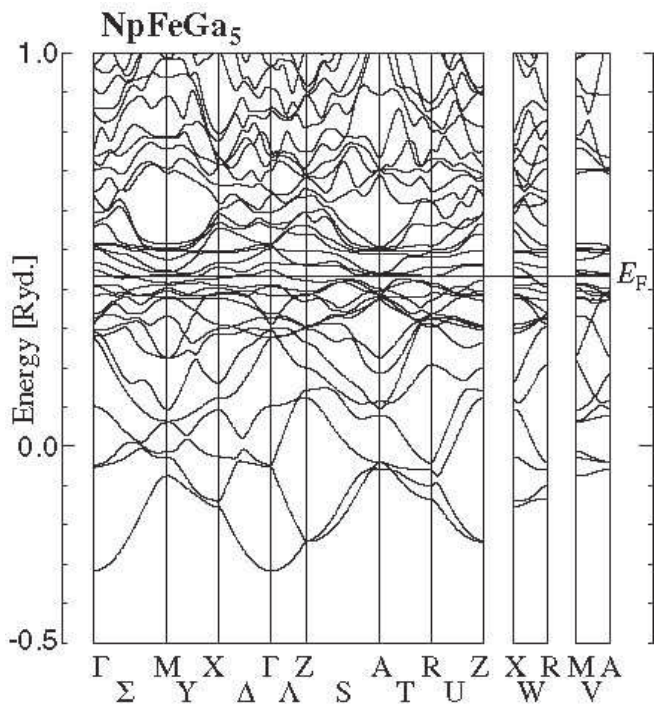}
\caption{Energy band structure calculated for NpFeGa$_5$ with
the self-consistent RLAPW method.}
\label{fi13}
\end{figure}

Finally, let us discuss the electronic properties of NpFeGa$_5$.
In Fig.~\ref{fi13}, the energy band structure along the symmetry axes
in the Brillouin zone is shown.
The Fermi energy $E_{\rm F}$ is located at $E_{\rm F}$=0.433 Ryd.
Again we observe that the electronic structure of NpFeGa$_5$ is 
similar to NpCoGa$_5$, except for the location of the Fermi level.
The number of the valence electrons in the APW sphere is
listed in Table~\ref{ta8}. 
There are 8.03 valence electrons outside the APW sphere in the
primitive cell and each Np APW sphere contains about 4.13 electrons
in the $f$ state.
The total density of states is calculated at $E_{\rm F}$ as
$N(E_{\rm F})=133.6$ states/Ryd.cell,
leading to $\gamma_{\rm band}$=23.1 mJ/K$^2 \cdot$mol.

\begin{figure}[t]
\includegraphics[width=1.0\linewidth]{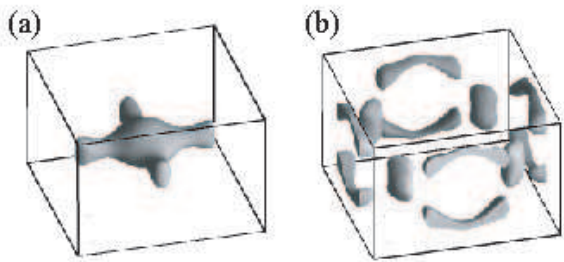}
\caption{Calculated Fermi surfaces of ${\rm NpFeGa_{5}}$ 
for (a) 15th band hole sheets and (b)16th band electron sheets.}
\label{fi14}
\end{figure}

We note that the lowest fourteen bands are fully occupied,
the next two bands are partially occupied,
and higher bands empty.
Then, 15th and 16th bands construct the Fermi surfaces,
as shown in Fig.~\ref{fi14}.
These Fermi-surface sheets are small in the size and closed in topology.
Note that there exists no open orbit on any sheets.
The numbers of carriers contained in these Fermi-surface sheets are
0.094 holes/cell and 0.094 electrons/cell in the 15th and 16th bands,
respectively.
The total number of holes is equal to that of electrons,
which means that NpFeGa$_5$ is a compensated metal.
Note that there are just 0.094 holes/cell and the compensating
number of electrons, indicating that NpFeGa$_5$ is a semimetal.

\begin{table}[t]
\caption{
The number of the valence electrons for NpFeGa$_5$
in the Np APW sphere, the Fe APW sphere,
and the Ga APW sphere partitioned into angular momenta.}
\label{ta8}
\begin{tabular}{lrrrr}
\multicolumn{1}{c}{ } & 
\multicolumn{1}{c}{$s$} & 
\multicolumn{1}{c}{$p$} & 
\multicolumn{1}{c}{$d$} & 
\multicolumn{1}{c}{$f$}   \\ \hline
Np & 0.32 & 6.04 & 1.79 & 4.13 \\
Fe & 0.40 & 0.38 & 6.40 & 0.01 \\
Ga($1c$) & 1.08 & 0.90 & 9.98 & 0.02 \\
Ga($4i$) & 3.85 & 2.92 & 39.79 & 0.06 \\
\end{tabular}
\end{table}

\subsubsection{Comment on NpTGa$_5$}

Here we discuss similarity and difference
among three Np-115 compounds.
For the purpose, let us examine the trend in the change of
$d$ electron number in transition metal atoms.
The numbers of electrons with the $d$ character contained in the
Fe, Co, and Ni APW spheres are 6.40, 7.49, and 8.39 in NpFeGa$_5$,
NpCoGa$_5$, NpNiGa$_5$, respectively,
indicating clearly that the number is increased by unity
among three Np-115 compounds.
Since one more $d$ electron is added to NpCoGa$_5$,
the Fermi level for NpNiGa$_5$ is shifted upward
in comparison with that of NpCoGa$_5$.

\begin{figure}[t]
\includegraphics[width=1.0\linewidth]{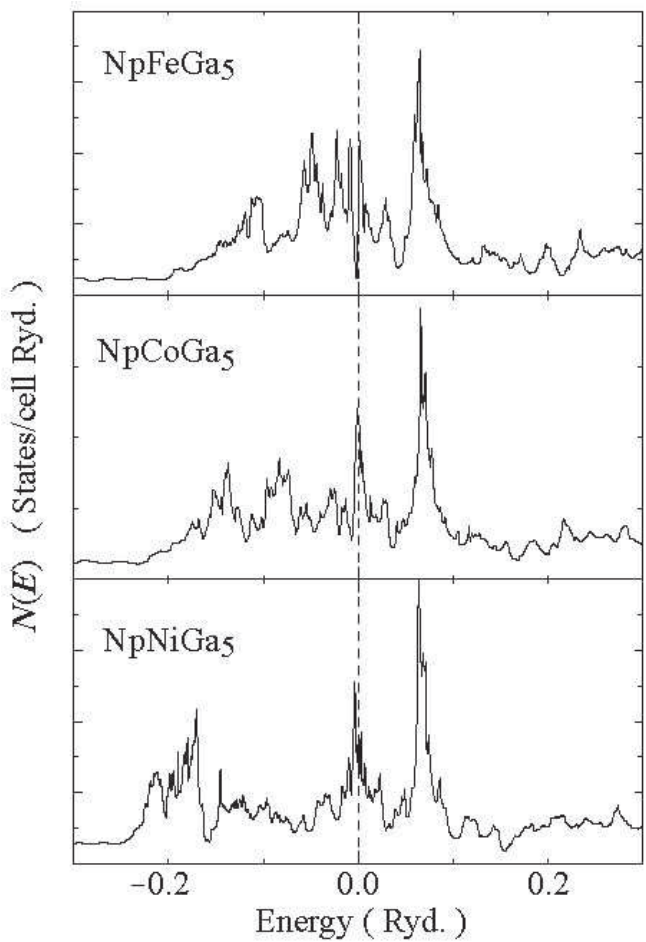}
\caption{Density of states for NpFeGa$_5$, NpCoGa$_5$ and 
NpNiGa$_5$. Dashed line indicates the Fermi energy, from which
the energy is measured.}
\label{fi15}
\end{figure}

To understand the change in the electronic properties among
three Np-115 compounds, it is useful to see the total
density of states (DOS), as shown in Fig.~\ref{fi15}.
We should note that there is no qualitative difference
in the energy-band structure among Np-115 compounds:
The $5f$ bands are split into two subbands by the spin-orbit
interaction and due to the hybridization between $5f$ and $4p$
electrons, finite DOS always appear at the Fermi level,
indicating that Np-115 compounds are metallic in our band-structure
calculations.
Note, however, that the peak structure in the DOS are located
just at or near the Fermi energy in common with three Np compounds.
This fact may be related to the easy appearance of magnetism
in Np-115 materials.
When the transition metal atom is substituted, the position of
$E_{\rm F}$ is a little bit shifted
upward with increasing $d$ electron in the order of
NpFeGa$_5$, NpCoGa$_5$, and NpNiGa$_5$.

\subsection{Result for AmCoGa$_5$}

\begin{figure}[t]
\includegraphics[width=1.0\linewidth]{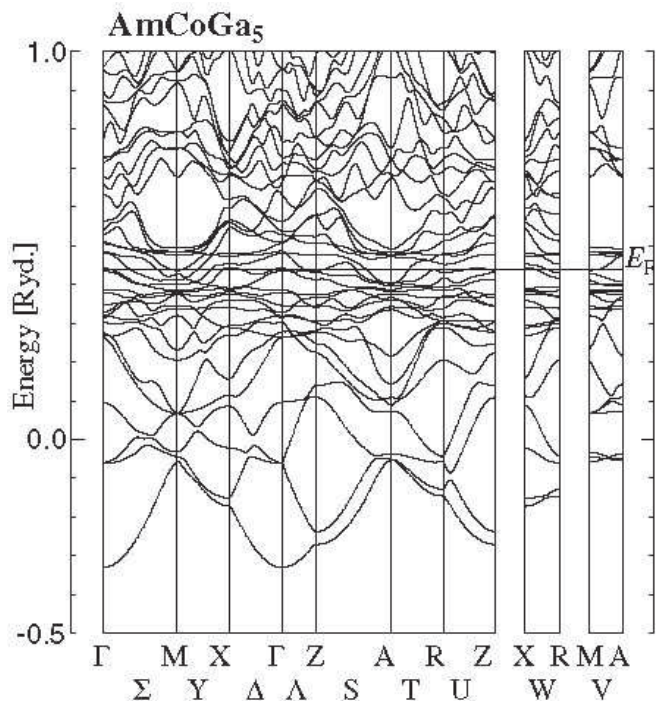}
\caption{Energy band structure for AmCoGa$_5$ calculated by
using the self-consistent RLAPW method.
$E_{\rm F}$ indicate the position of the Fermi level.}
\label{fi15a}
\end{figure}

Now we show our calculated result of the energy-band structure
of AmCoGa$_5$ within the framework of the itinerant-electron model
for the 5$f$ electrons.
In Fig.~\ref{fi15a}, we depict the energy band structure
along the symmetry axes in the Brillouin zone
in the energy region from $-0.5$ Ryd. to 1.0 Ryd.
We do not show the three Am $6p$ and twenty-five Ga $3d$ bands
in the energy range between $-1.0$ Ryd. and $-0.6$ Ryd.
The Fermi level $E_{\rm F}$ is located at 0.437 Ryd. and
in the vicinity of $E_{\rm F}$, there occurs a hybridization between
the Am 5$f$ and Ga 4$p$ states.
The number of the valence electrons in the APW sphere is partitioned
into the angular momenta, as listed in Table \ref{ta8a}.
There are 8.22 valence electrons outside the APW sphere
in the primitive cell.
The total density of states at $E_{\rm F}$ is evaluated as 
$N(E_{\rm F})$=62.8 states/Ryd.cell.
By using this value, the theoretical specific heat coefficient
$\gamma_{\rm band}$ is estimated as 10.9 mJ/K$^2 \cdot$mol.

\begin{table}[t]
\caption{The number of valence electrons for AmCoGa$_5$
in the Am APW sphere, the Co APW sphere,
and the Ga APW sphere partitioned into angular momenta.}
\label{ta8a}
\begin{tabular}{lrrrr}
\multicolumn{1}{c}{ } &
\multicolumn{1}{c}{$s$} &
\multicolumn{1}{c}{$p$} &
\multicolumn{1}{c}{$d$} &
\multicolumn{1}{c}{$f$} \\ \hline
Am & 0.39 & 6.16 & 1.68 & 6.39 \\
Co & 0.43 & 0.42 & 7.45 & 0.01 \\
Ga($1c$) & 0.95 & 0.67 & 9.92 & 0.01 \\
Ga($4i$) & 3.73 & 2.77 & 39.75 & 0.05 \\
\end{tabular}
\end{table}

\begin{figure}[t]
\includegraphics[width=1.0\linewidth]{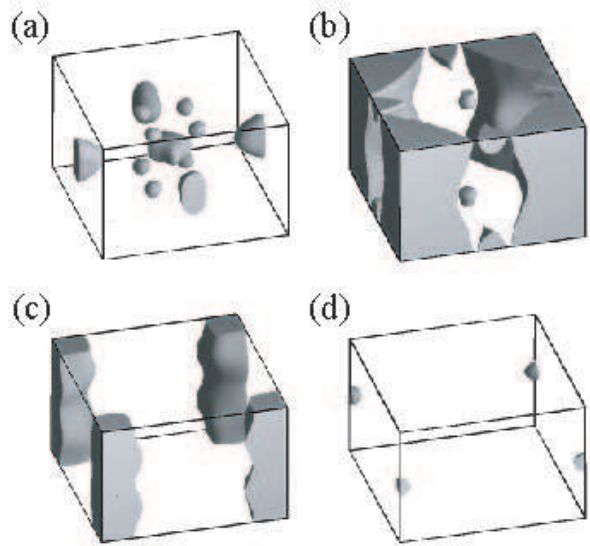}
\caption{Calculated Fermi surfaces of AmCoGa$_5$ for
(a) 16th band hole sheets, (b) 17th band hole sheets, 
(c) 18th band electron sheets, and (d) 19th band electron sheets.}
\label{fi15b}
\end{figure}

In Fig.~\ref{fi15a}, the lowest fifteen bands are fully occupied.
The next four bands are partially occupied, while higher bands are
empty. Namely, 16th, 17th, 18th, and 19th bands crossing
the Fermi level construct the hole or electron sheet of the Fermi
surface, as shown in Fig.~\ref{fi15b}.
As shown in Fig.~\ref{fi15b}(a),
the Fermi surface from the 16th band consists of
two equivalent small hole sheets centered 
at the X points and one hole sheet centered at the $\Gamma$ point.
It consist of eight small closed sheets which have the mirror-inversion 
symmetry with respect of $\{100\}$, $\{010\}$, and $\{001\}$ planes
lying between them.
The 17th band has three kinds of sheets, as shown in Fig.~\ref{fi15b}(b).
One is a set of two equivalent electron-like pockets,
each of which is centered at the R point.
Another is electron pocket centered at the Z point.
There is also a large cylindrical electron sheet
centered at the M point, which characterizes
the two-dimensional Fermi surface.
The 18th band has a slender cylindrical electron sheet
which is also centered at the M point,
as shown in Fig.~\ref{fi15b}(c).
The 19th band constructs a very small electron sheet,
which is centered at the M point, as shown in Fig.~\ref{fi15b}(d).
The number of carriers contained in these Fermi-surface sheets are
0.048 holes/cell, 0.785 electrons/cell, 0.272 electrons/cell, and
0.0004 electrons/cell in the 16th, 17th, 18th, and 19th bands,
respectively.
The total number of holes is not equal to that of electrons,
indicating that ${\rm AmCoGa_{5}}$ is an uncompensated metal.

We remark that the Fermi surfaces of AmCoGa$_{5}$ seem
to be similar to those of PuCoGa$_{5}$, since
we can see the multiple two-dimensional cylindrical
Fermi-surface sheets with a constricted part around at M point.
For a better understanding of the Fermi surface, we attempt to
explain the relationship between AmCoGa$_{5}$ and PuCoGa$_{5}$.
On the basis of the itinerant 5$f$ electron model, AmCoGa$_{5}$ 
has one more Bloch electron per primitive cell in comparison with
PuCoGa$_{5}$.
Therefore, in AmCoGa$_{5}$, the Fermi level shift upward,
indicating that relatively the position of 5$f$ bands move
downward.
As a result, we find that the 15th-band hole sphere of PuCoGa$_5$
disappears and the size of the 16th-band hole sheet becomes smaller.
On the other hand, the sizes of the 17th- and 18th-band cylindrical
electron sheets become large.
Note that in AmCoGa$_5$, a small electron sheet appears
in the 19th band.

However, in the recent band-structure calculation result,
Opahle {\it et al.} have found that the Fermi surface structure of
AmCoGa$_5$ is different from that of PuCoGa$_5$.\cite{Opahle2}
Although the figures for band structure and Fermi surface of
AmCoGa$_5$ were not shown in Ref.~38,
their results for PuCoGa$_5$ have been found to be quite similar
to ours and thus, their Fermi surfaces of AmCoGa$_5$ are considered
to be different from ours.
One reason for this discrepancy is the treatment of 5f electrons.
Namely, we have treated itinerant $5f$ electrons,
while Opahle {\it et al.} considered localized ones.
In general, the tendency of localization of $5f$ electron in Am
is stronger than that in Pu, but in actual compounds, it is
difficult to determine which is the better approximation,
localized or itinerant picture.
It will be necessary to clarify which picture well explains
the experimental results consistently.

Finally, let us comment on the effect of initial electron
configurations in actinide atoms.
In the present calculations, as we have mentioned in Sec.~2,
we have considered the electron configuration of
([Rn]5$f^6$6$d^1$7$s^2$) for Am atom.
However, this configuration is not uniquely determined.
For instance, it may be possible to start the calculation by using
([Rn]5$f^7$6$d^0$7$s^2$).
Also for Np- and Pu-115 compounds, we can consider other
electron configurations such as
Np([Rn]5f$^5$6$d^0$7$s^2$) and Pu([Rn]5$f^6$6$d^0$7$s^2$).
In order to finalize the Fermi-surface structure of actinide 115
materials, it is necessary to perform carefully further calculations
by changing the initial electron configurations for actinide ions.
It is an important future task.

\section{Discussion}

\begin{figure}[t]
\includegraphics[width=1.0\linewidth]{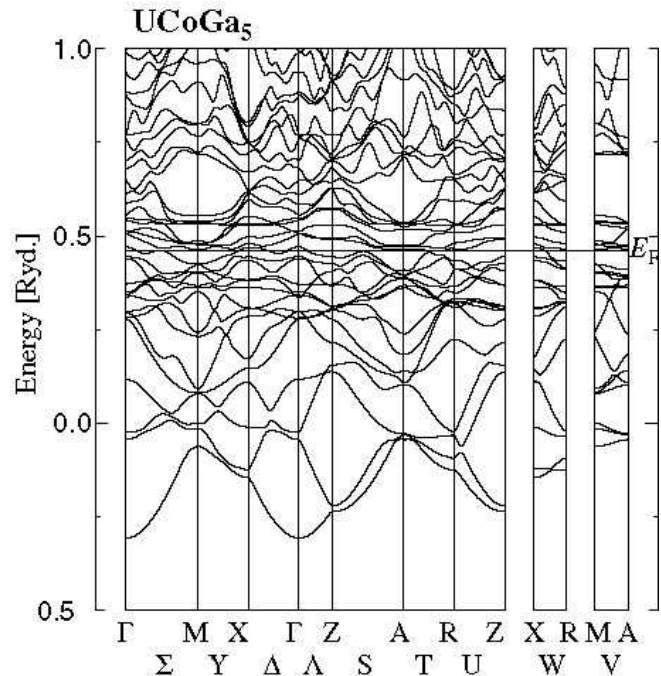}
\caption{Energy band structure calculated for UCoGa$_5$ with
the self-consistent RLAPW method.}
\label{fi16}
\end{figure}

In the previous section, we have shown the band-structure
calculation results for PuTGa$_5$, NpTGa$_5$, and AmCoGa$_5$.
In order to obtain deep insight into the electronic properties of
these 115 compounds, it is useful to compare the results with
the electronic energy band structure of UCoGa$_5$,
as shown in Fig.~\ref{fi16}.\cite{Maehira3,Maehira4}
Note that three U $6p$ and twenty-five Ga $3d$ bands are omitted.
The Fermi energy $E_{\rm F}$ is located at $E_{\rm F}$=0.461 Ryd.
The total density of states at $E_{\rm F}$ of UCoGa$_5$ is
calculated as $N(E_{\rm F})$=48.4 states/Ryd.cell,
which corresponds to $\gamma_{\rm band}$=8.4 mJ/K$^2 \cdot$mol.

\begin{figure}[t]
\includegraphics[width=1.0\linewidth]{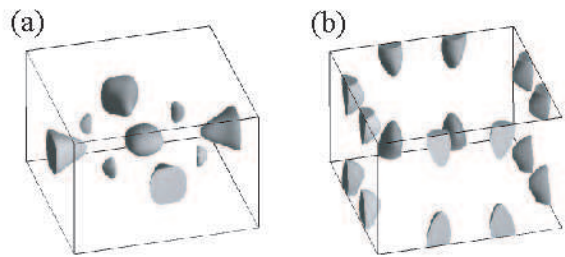}
\caption{Calculated Fermi surfaces of ${\rm UCoGa_{5}}$
for (a) 15th band hole sheets and (b)16th band electron sheets.}
\label{fi17}
\end{figure}

In UCoGa$_5$, the 15th and 16th bands form the Fermi surfaces,
as shown in Fig.~\ref{fi17}.
The Fermi surfaces from the 15th band have
one sheet centered at the $\Gamma$ point,
two equivalent sheets centered at the X points,
and the sheets across the S axis.
Figure \ref{fi17}(b) shows a set of the sixteen electron sheets
of the Fermi surfaces in the 16th band.
Each electron sheet across the T axis looks like a cushion.
The total number of holes is equal to that of electrons,
which means that UCoGa$_5$ is a compensated metal.

Except for details, we can observe that the sheets of the Fermi surface
of UCoGa$_5$ with small size and closed topology are similar to
those of NpFeGa$_5$ (see Fig.~\ref{fi14}).
We also remark that
NpCoGa$_5$ is considered to be UCoGa$_5$ plus one more $f$-electron
from the viewpoint of the Fermi surface topology,
since the small-pocket parts are the remnants of UCoGa$_5$
and the large-volume Fermi surface contains one additional electron.
Interestingly enough, the Fermi surfaces of UNiGa$_5$
\cite{U115-2} are quite similar to Fig.~\ref{fi10},
since both of UNiGa$_5$ and NpCoGa$_5$
are regarded as UCoGa$_5$ plus one more electron,
if we simply ignore the difference in the original character,
$d$ or $f$, of the additional one electron.
Since NpCoGa$_5$ is UCoGa$_5$ plus one $f$-electron
and NpFeGa$_5$ is regarded as NpCoGa$_5$ minus one $d$-electron,
the Fermi surfaces of UCoGa$_5$ are similar to NpFeGa$_5$.

Based on the discussion on the number of $d$ and $f$ electrons,
we also point out that the Fermi surfaces
of NpNiGa$_5$ should be similar to those of PuCoGa$_5$,
as shown in Figs.~\ref{fi3} and \ref{fi12}.
However, it is clearly observed that two-dimensionality in the
Fermi surfaces of NpNiGa$_5$ becomes
worse compared with those of PuCoGa$_5$.
This may be related to the reason why NpNiGa$_5$ is $not$ superconducting,
in spite of the same group with PuCoGa$_5$
in the electron number discussion.

The above phenomenology on the $d$- and $f$-electron numbers seems
to suggest that a rigid-band picture works for actinide 115 compounds.
This point can be partly validated due to the fact that
the band structure of these compounds is basically determined
in common by hybridization between broad $p$ bands and 
narrow $f$ bands in the vicinity of the Fermi level.
Based on the rigid-band picture, the electron band structure itself
is not changed significantly, even when we change
actinide and/or transition metal ions.
Thus, just by counting the valence electrons of actinide and transition
metal ions, we can easily deduce the Fermi-surface topology,
starting with the results of UCoGa$_5$.

\begin{figure}[t]
\includegraphics[width=0.6\linewidth]{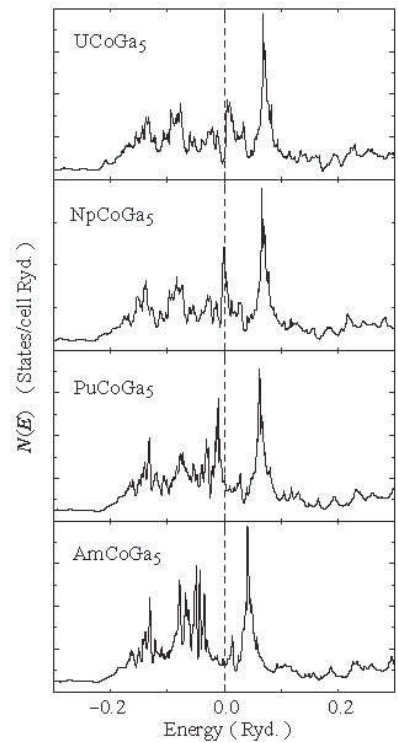}
\caption{Density of states for UCoGa$_5$, NpCoGa$_5$, PuCoGa$_5$ 
and AmCoGa$_5$. Dashed line indicates the Fermi energy.}
\label{fi18}
\end{figure}

It is interesting to compare the total DOS calculated for AnCoGa$_5$
(An=U, Np, Pu, and Am), as shown in Fig.~\ref{fi18}.
We again see that the structures of DOS are qualitatively in common
with actinide 115 compounds except for details,
when we change the actinide ions.
In UCoGa$_5$, the Fermi energy is located in a valley,
leading to the small DOS at $E_{\rm F}$,
consistent with a semi-metal behavior.
On the other hand, for NpCoGa$_5$, as mentioned in the previous section,
the peak is located just at the Fermi energy.
After including the effect of electron correlation,
magnetic transition may easily occur in this case.
For PuCoGa$_5$ and AmCoGa$_5$, we observe moderate values of
the DOS at the Fermi energy, consistent with the paramagnetic
metallic phase.

It is noted that the numbers of electrons with the $f$ symmetry
contained in the actinide APW sphere are, respectively,
3.2, 4.2, 5.2, and 6.4
for UCoGa$_5$, NpCoGa$_5$, PuCoGa$_5$, and AmCoGa$_5$,
with increasing by unity
as the atomic number of actinide atom increases.
This fact seems to suggest that electrons are supplied to
the unoccupied 5$f$ bands of UCoGa$_5$.
Namely, the change in the Fermi surface structures
among actinide 115 compounds is basically understood by
an upward shift of the Fermi level on the rigid electronic bands.
The same change occurs between PuCoGa$_5$ and AmCoGa$_5$.
Note, however, that it is also important to specify
the difference among 115 compounds.
Especially, the change of the magnetic structure in Np-115 compounds
as well as the appearance of moment at Fe ion in NpFeGa$_5$
\cite{Honda,Yamamoto,Homma,Metoki} is not
clarified by the present band-structure calculations.
In order to understand these points, 
it is necessary to calculate not only the total DOS but also
the partial DOS for $d$- and $f$-electron component
within our band-calculation program.
It is one of future problems.

Finally, let us discuss the dHvA experimental results on
NpNiGa$_5$\cite{Aoki-Ni}, NpCoGa$_5$,\cite{Aoki-Co2} and
NpRhGa$_5$,\cite{Aoki-Rh}
in comparison with our theoretical Fermi surfaces of Np-115.
For NpNiGa$_5$, Aoki {\it et al.} have observed several sets of
the dHvA frequency branches in the region of the order 10$^{7}$ Oe,
which have the cyclotron effective masses from 1.8$m_{0}$ to
4.9$m_{0}$ in the magnetic-field direction [100] and [001].
Here $m_{0}$ is the rest mass of a free electron.
It is remarkable that the effective masses greater than the 
free-electron mass have been observed, because it implies that
the 5$f$ electrons may be itinerant in NpNiGa$_5$.

For NpCoGa$_5$, a couple of cylindrical sheets of Fermi surfaces
with large volume have been detected, while in our band-structure
calculations, one cylindrical sheet is obtained, as shown in
Fig.~\ref{fi10}.
Namely, the dHvA results are not explained by the band calculation
in the paramagnetic state.
A possible explanation for this discrepancy is to consider
the folding of Fermi surfaces in the magnetic Brillouin zone,
since NpCoGa$_5$ exhibits the A-type AF structure
with a N\'eel temperature $T_{\rm N}$=47K.
If the magnetic unit cell is elongated along the [001] direction
and doubled with respect to the chemical unit cell
as observed in UPtGa$_{5}$,\cite{U115-8}
a quasi-two dimensional Fermi surface 
is more likely to appear due to the flat magnetic Brillouin zone.

However, in actual dHvA experiments, due to the applied magnetic
field, the system is not AF, but ferromagnetic.
Thus, the above explanation cannot be simply applied to this case.
In addition, quite recently, the dHvA experiments have been also
performed for NpRhGa$_5$, which is also A-type antiferromagnet.
In this material, four cylindrical sheets of Fermi surfaces
have been reported,\cite{Aoki-Rh}
consistent with the folding of Fermi surfaces
of NpCoGa$_5$ in the magnetic Brillouin zone.
Thus, it is experimentally confirmed that a couple of cylindrical
sheets of Fermi surfaces exist in NpCoGa$_5$.

It may be true that the rigid-band picture cannot explain
universally the band structure of actinide 115 compounds,
especially of Np-115 materials.
However, a simple tight-binding approximation based on the
$j$-$j$ coupling scheme can reproduce a couple of cylindrical
Fermi surfaces,\cite{Onishi} by assuming that two of four $5f$
electrons in Np ion are active.
In order to understand the dHvA experimental results
on the Fermi surfaces, it is a correct direction
to improve the band-structure calculations by considering
the difference in the degree of itinerancy among $5f$ electrons.
In fact, it has been suggested that a couple of cylindrical
Fermi-surface sheets are obtained in the spin-orbital polarized
band-structure calculation.\cite{Yamagami2}
Another way is an application of the full-potential method
to the fully relativistic band-structure calculations.
Such an extension of the band-calculation method
is one of important future tasks.

%
%
\section{Summary}

In this paper, we have applied the RLAPW method to the self-consistent
calculation of the electronic structure for PuTGa$_5$, NpTGa$_5$,
and AmCoGa$_5$ on the basis of the itinerant $5f$ electron picture.
It has been found that a hybridization between the $5f$ and
Ga $4p$ states occurs in the vicinity of $E_{\rm F}$.
We have calculated the dHvA frequencies for PuCoGa$_5$
for future experiments.
The similarity in the Fermi surface structure among actinide 115
compounds has been found to be understood based on the rigid-band
picture, while the theoretical Fermi surfaces of NpCoGa$_5$
are different from ones in the recent dHvA experimental results.
For the purpose to understand this point,
it is a challenging future problem to improve the
band-calculation technique.

\section*{Acknowledgements}

The authors thank F. Wastin for useful information
on transuranium 115 compounds.
We also thank D. Aoki, Y. Haga, Y. Homma, F. Honda, S. Ikeda,
K. Kaneko, K. Kubo, T. D. Matsuda, N. Metoki, A. Nakamura,
H. Onishi, Y. \=Onuki, H. Sakai, Y. Shiokawa, R. E. Walstedt,
H. Yamagami, E. Yamamoto, and H. Yasuoka for fruitful discussions.
One of the authors (T. H.) has been suppprted by Grant-in-Aids
for Scientific Research (No.~14740219) of Japan Society for
the Promotion of Science and
for Scientific Research in Priority Area ``Skutterudites''
(No.~16037217) of the Ministry of Education, Culture, Sports,
Science, and Technology of Japan.
The computation of this work has been done using the facilities
of Japan Atomic Energy Research Institute.


\end{document}